\documentclass[preprint2]{aastex}
\usepackage{emulateapj5}
\usepackage{epsfig}
\usepackage{onecolfloat5}
\begin{document}

\def\deg{^{\circ}}

\newcommand{\psrae}{\mbox{J0737$-$3039A}}
\newcommand{\psra}{\mbox{J0737$-$3039A }}
\newcommand{\psrbe}{\mbox{J0737$-$3039B}}
\newcommand{\psrb}{\mbox{J0737$-$3039B }}
\newcommand{\be}{\begin{eqnarray}}
\newcommand{\ee}{\end{eqnarray}}

\shorttitle{Modulation of the emission from B by A}

\twocolumn[
\title{The Double Pulsar System J0737--3039: \\ Modulation of the radio emission from B by radiation from A}
\author{M.\ A.\ McLaughlin\altaffilmark{1},  M.\ Kramer\altaffilmark{1}, A. G. Lyne\altaffilmark{1},  D.\ R.\ Lorimer\altaffilmark{1}, \\ I. H. Stairs\altaffilmark{2}, A. Possenti\altaffilmark{3}, R.\ N.\ Manchester\altaffilmark{4}, P. C. C. Freire\altaffilmark{5}, \\ B. C. Joshi\altaffilmark{6}, M.\ Burgay\altaffilmark{3}, F. Camilo\altaffilmark{7} \& N.\ D'Amico\altaffilmark{8}}

\begin{abstract}

We have analyzed single pulses from PSR~\psrbe, the 2.8-s pulsar in the
recently discovered double pulsar system, using data taken with the Green
Bank Telescope at 820 and 1400 MHz. We report the detection of features
similar to drifting subpulses, detectable over only a fraction of the
pulse window, with a fluctuation frequency of 0.196 cycles/period. This is
exactly the beat frequency between the periods of the two pulsars.  In
addition, the drifting features have a separation within a given pulse of
23~ms, equal to the pulse period of A.  These features are therefore due
to the direct influence of \psrae's 44-Hz electromagnetic radiation on
\psrbe's magnetosphere. We only detect them over a small range of orbital
phases, when the radiation from the recycled pulsar \psra meets our line
of sight to \psrb from the side.

\end{abstract}

]

\altaffiltext{1}{Jodrell Bank Observatory, University of Manchester, Macclesfield, Cheshire, SK11 9DL, UK}
\altaffiltext{2}{Dept. of Physics and Astronomy, University of British
Columbia, 6224 Agricultural Road, Vancouver, BC V6T 1Z1 Canada}
\altaffiltext{3}{INAF - Osservatorio Astronomico di Cagliari, Loc. Poggio dei
Pini, Strada 54, 09012 Capoterra, Italy}
\altaffiltext{4}{Australia Telescope National Facility -- CSIRO,
P.O. Box 76, Epping NSW 1710, Australia}
\altaffiltext{5}{NAIC, Arecibo Observatory, HC03 Box 53995, PR 00612, USA}
\altaffiltext{6}{National Centre for Astrophysics, P.O. Bag 3, Ganeshkhind, Pune 411007, India
}
\altaffiltext{7}{Columbia Astrophysics Laboratory, Columbia University, 550 West 120th Street, New York, NY 10027, USA}
\altaffiltext{8}{Universit\`a degli Studi di
Cagliari, Dipartimento di Fisica, SP Monserrato-Sestu km
0.7, 09042 Monserrato, Italy}

\section{Introduction} \label{sec:intro}

The discovery of the first double pulsar binary system, J0737--3039
(Burgay et al.~2003; Lyne et al.~2004) presents a unique opportunity to
study new aspects of relativistic gravity and relativistic plasma physics.
The two pulsars in this system, \psra and \psrb (hereafter simply ``A''
and ``B'') have periods of $P_A = 23$~ms and $P_B = 2.8$~s and are in a
2.4-hr mildly-eccentric orbit. Of particular interest here is the fact
that, because the rate of spin-down energy loss from A is $\sim$3600 times
greater than that from B, the radiation from A is expected to have a
significant impact on B.  In fact, at the light cylinder radius of B, the
energy density of the 44-Hz radiation from A is about two orders of
magnitude greater than that of the 0.36-Hz radiation from B.

Aside from a $\sim$ 30-s duration eclipse (Lyne et al.~2004; Kaspi et
al.~2004), the emission from the more energetic pulsar A is very steady
and is consistent with that seen from other recycled pulsars. The emission
from pulsar B, however, exhibits extreme variations in its flux density
over a single orbit. At two orbital phases near the inferior conjunction
of B (i.e. when B is closest to Earth), bright single pulses are
detectable and can be studied in detail while at some other orbital
phases, no radio emission is detectable at all.  To explain these
variations, Jenet \& Ransom (2004) proposed a geometric model in which B
is stimulated to emit when it is illuminated by A's hollow-coned radio
beam. Zhang \& Loeb (2004)  suggest that this occurs as pairs from A's
wind flow into the open field line region of B and emit curvature
radiation. Alternatively, it seems possible that the charges in A's wind
short out the currents in the magnetosphere of B at certain orbital
phases. While these models are plausible, no direct proof of an
interaction between the two pulsars has so far been found.

In this {\it Letter}, we present the first direct observational evidence
for the impact of A's electromagnetic radiation on B from an analysis of
single pulses from B. This has revealed a phenomenon similar to the
drifting subpulses detected from many radio pulsars (e.g. Drake \& Craft 1969,
Backer 1973, Taylor, Manchester \& Huguenin 1975). However, in this
case, the features are only detected at certain orbital phases. In \S2 we
describe our observations and analysis and characterize the drifting
behavior. In \S3, we show that this phenomenon is due to the direct
influence of A's electromagnetic radiation on B's magnetosphere and
explain why these features are only seen at certain orbital phases.

\section{Observations and Analysis} \label{sec:obsandresults}

We have analyzed the now publicly-available exploratory-time observations
of the double pulsar system obtained with the 100-m Green Bank Telescope
(GBT) in 2003 December and 2004 January and using receivers at 427, 820
and 1400 MHz.  The 427-MHz and 820-MHz data discussed here were acquired
with the GBT Spectrometer SPIGOT card with sampling times of 81.92~$\mu$s
and 40.96~$\mu$s at those two frequencies, respectively.  At both
frequencies, 1024 synthesized frequency channels covered a 50-MHz
bandwidth. The 1400-MHz data were acquired using the Berkeley-Caltech
Pulsar Machine (BCPM) using a sampling interval of 72~$\mu$s on each of 96
channels covering a 96-MHz bandwidth. For further details of the data
acquisition systems, see Ransom et al.~(2004) and references therein.

\begin{figure}[t]
\begin{center}
\epsfig{file=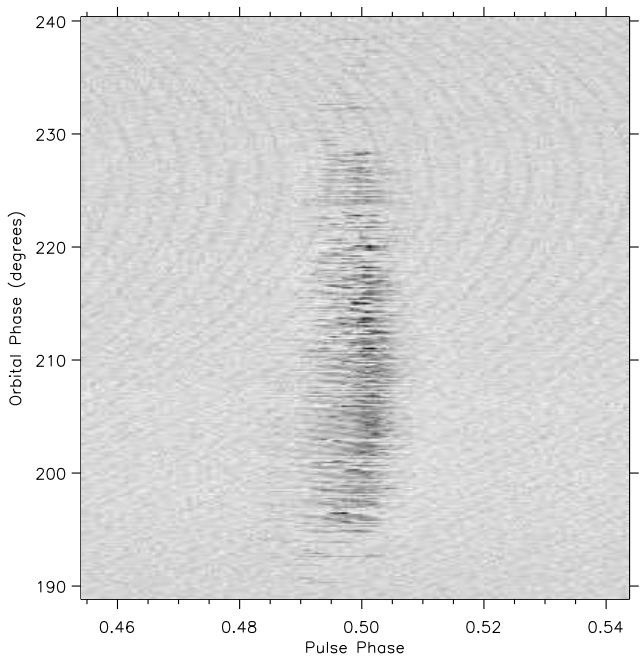,height=9cm}
\caption{
\label{fig:single1}
Single pulses of B at 820 MHz for orbital phases 190 -- 240$\deg$ (i.e.
403 pulses) on MJD 52997. Only 10\% of the pulse period of B is shown.
Drifting features are present through most of these data, but are
particularly obvious from orbital phases $\sim$ 195 -- 210$\deg$. Single
pulses from both components of A may also be seen in the background, most
clearly at orbital phase $\sim$ 225$\deg$, where differential Doppler
shifts from the orbital motion result in harmonically related apparent
pulse periods. An expanded view of the drifting region is provided in
Fig.~5.
}
\end{center}
\end{figure}
\begin{figure}[h]
\begin{center}
\epsfig{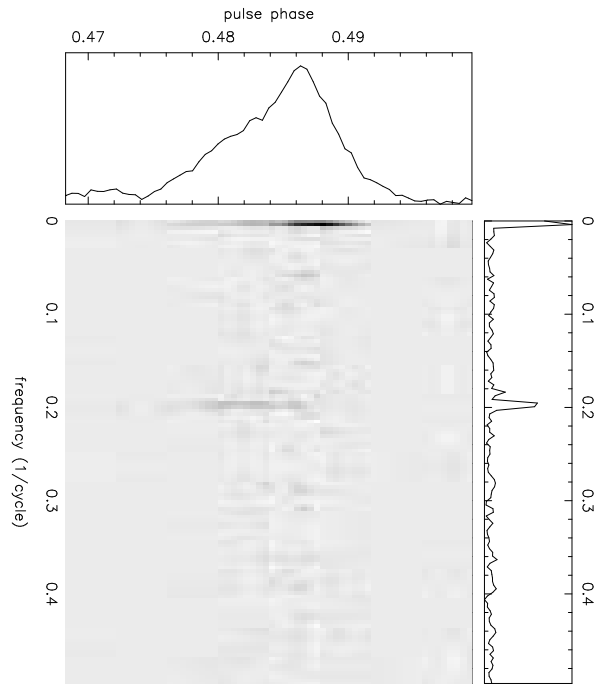}
\caption{
\label{fig:drift1}
Upper: Integrated pulse profile of B for orbital phases 190 -- 240$\deg$.
Only 64 of 2048 original phase bins are shown. Middle: Grey-scale plot of
the fluctuation spectrum of B, with the x-axis corresponding to pulse
longitude and the y-axis to fluctuation frequency. Overlapping spectra of
256-pulse sequences have been computed. A feature with frequency of 0.196
cycles/period near the front half of the pulse profile is obvious. The
low-frequency feature is due to the large change in the flux of B over this
orbital phase range. Right:
Spectrum summed over pulse phases 0.478 -- 0.486, where the feature at
0.196 cycles/period
is brightest.
}
\end{center}
\end{figure}

The GBT data were dedispersed and folded using freely available software
tools (Lorimer 2001) assuming the nominal dispersion measure
(48.914~cm$^{-3}$~pc; Burgay et al.~2003) and using an ephemeris for
pulsar B from Lyne et al. (2004).  As reported in that paper, and in
Ramachandran et al. (2004), we find that the B pulsar is brightest at
orbital phases 195--230$\deg$ and 260--300$\deg$, where we define orbital
phase as the longitude from ascending node (i.e. the sum of the longitude
of periastron and the true anomaly).  In Fig.~1, we
present a sequence of single pulses from B at 820 MHz between orbital
phases 190 -- 240$\deg$.  At these orbital phases, the pulses of B are
characterized by a strong trailing pulse component with a weaker leading
component. In Fig.~1, single pulses from A are also
apparent.  The single pulses from A are very stable in flux, show no
obvious drifting behavior and are typical of those seen in other recycled
pulsars (e.g. Jenet, Anderson \& Prince 2001, Edwards \& Stappers 2003),
although the number of millisecond pulsars from which single pulses have
been detected is quite small.

A striking effect seen in Fig.~1, which was not reported
in an earlier analysis of these data by Ramachandran et al.~(2004), is a
drifting phenomenon occurring in the single pulses of B in the orbital
phase range $\sim$ 195 -- 215$\deg$.  The drifting features are confined
to the leading component of the pulse profile and are quite narrow, with a
width of $\sim~0.1\%~P_B$. In Fig.~2 we present a
longitude-resolved fluctuation spectrum, calculated as described in
Backer~(1973), for this pulse sequence. These spectra reveal the
frequencies at which the intensity of a pulsar signal fluctuates for a
range of pulse longitudes.  Because of the short amount of time over which
the single pulses from B are detectable, only 384 pulses were used for
this analysis. The fluctuation spectrum was calculated using two
overlapping spectra of 256-pulse sequences, for a spectral resolution of
(1/256$P_B$). We detect a strong feature at $0.196^{+0.005}_{-0.002}$
cycles/period (i.e. a modulation period $P_3 = 5.10 P_B$). The subpulse
separation within a given pulse, $P_2 = 23 \pm 1$ ms, is equal, within the
uncertainties, to the period of pulsar A. We see similar drifting behavior
at the same orbital phases and with exactly the same fluctuation frequency
one orbit later in the same 820 MHz dataset and in data taken with the
BCPM at 1400~MHz two weeks prior to the 820 MHz dataset. We also detect
the same drifting in 427-MHz data taken with the SPIGOT card but the lower
signal-to-noise ratio of single pulses (partly due to the large amount of
radio frequency interference) does not allow us to compute sensitive
fluctuation spectra. We detect the same drifting features very weakly in
data taken with the Parkes 64-m telescope in 2003, but the signal-to-noise
of those data is too poor to permit any detailed analysis.

\begin{figure}[t]
\begin{center}
\epsfig{file=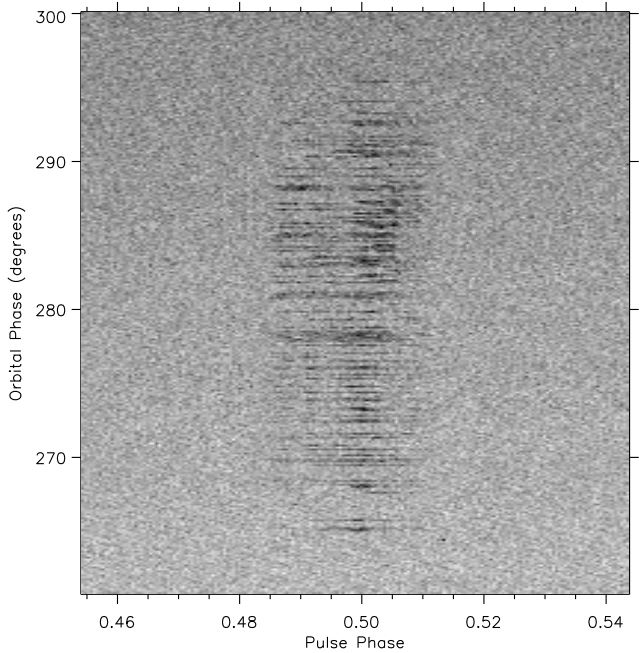,height=9cm}
\caption{
\label{fig:single2}
Single pulses of B at 820 MHz for orbital phases 260 -- 300$\deg$ (i.e.
296 pulses) on MJD 52997. As in Fig.~1, only 10\% of the
pulse period of B is shown. No drifting features are obvious, though the
single pulses do show ordered structure which may be influenced in some way
by A. Further studies of the single pulses in this orbital phase range are underway.}
\end{center}
\end{figure}

In Fig.~3, we show single pulses of B at orbital phases
260 -- 300$\deg$. The pulses in this orbital phase range are somewhat
weaker than those shown in Fig.~1. The profile is
generally broader, with a stronger leading component. No drifting is seen
at these orbital phases. In Fig.~4, we show a fluctuation
spectrum for the single-pulse sequence shown in Fig.~3.
Because of the shorter amount of time that single pulses are detectable in
this phase range, only 192 pulses were used for the analysis, with
overlapping 128-pulse spectra computed. No features indicative of drifting
are obvious for this range of orbital phases, although there is a larger
modulation index on the outer edges of the pulse.

\section{Discussion} \label{sec:discussion}

At first sight, the drifting we see in the single pulses of pulsar B is
similar to the drifting subpulses that have been detected from a number of
``normal'' radio pulsars. In the following discussion, however, we
demonstrate that, rather than being intrinsic to B's emission process, the
drifting features are a direct result of the impact of A's 44-Hz
electromagnetic radiation on B.

The ratio of the intrinsic barycentric rotation period of B to that of A
is 122.182. Because the periods are incommensurate, any influence of A's
electromagnetic radiation on B's magnetosphere will manifest itself as a
drifting behavior caused by the beating of the two periodicities.
Using the intrinsic periods of the two pulsars and their orbital
elements, we can accurately predict the frequency of A's pulsed radiation
as seen by B. We simply compute the apparent period of A as seen by B
given the barycentric period of A and the changing propagation time of A's
pulses due to the to the varying separation of the two pulsars, arising   
from the orbital eccentricity. We assume that the signal travels at the
velocity of light.  
At orbital phases 195$\deg$ -- 210$\deg$, where the drifting is
most obvious, the beat frequency between A and B varies from 0.200 to
0.196 cycles/period, exactly matching the detected feature at
$0.196^{+0.005}_{-0.002}$ cycles/period. Indeed, in Fig.~5
we show an expanded view of the single pulses from Fig.~1
for the range of orbital phases of interest with the predicted phases at
the center of mass of B of a signal with A's periodicity. It is clear that
the drifting pattern closely follows the predicted variation in phase.
This, and the fact that the separation of successive features within a
given pulse is equal to $P_A$, makes a convincing case that the drifting
is due to the direct influence of a signal with A's periodicity on the B
pulse emission mechanism. The fact that the observed modulation is at
44~Hz with only a single pulse in each 23-ms period suggests that it is
not the beamed radiation from A, which has two pulses per period, that
excites the B emission. On these as well as energetic grounds, we conclude
that the observed modulation is due to the influence of the 44-Hz magnetic
dipole radiation on the magnetosphere of B. Furthermore, the modulation is
caused by the electromagnetic field itself, rather than its intensity or
pressure, which have an 88-Hz periodicity\footnote{The power in the
fluctuation spectrum at 0.392 cycles/period is a factor of $\sim$ four
smaller than the power at 0.196 cycles/period.}. We note that this result
provides observational support to the idea that, close to the pulsar, most
of the spin-down energy is carried by the Poynting flux of the
magnetic-dipole radiation rather than by energetic particles (e.g. Michel
1982).

\begin{figure}[t]
\begin{center}
\epsfig{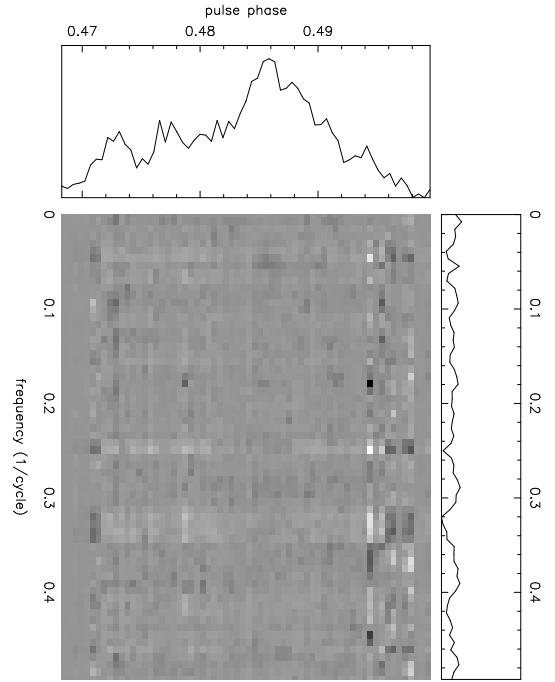}
\caption{
\label{fig:drift2}
As in Fig.~2, but for orbital phases 260 -- 300$\deg$. No
features indicative of drifting behavior are seen in the fluctuation
spectrum.  Because the flux of B is more stable at these orbital phases,
there are no obvious low-frequency features. Given the model discussed in
Section~\ref{sec:discussion}, fluctuation frequencies of 0.180 to 0.166~cycles/period would be
expected.}
\end{center}
\end{figure}

We can think of two different ways in which the electromagnetic radiation
from A could influence the processes in B's magnetosphere. It is possible
that the electric component of A's radiation field accelerates the
radiating particles in the magnetosphere of B. Alternatively, the magnetic
component of A's radiation field could modulate the magnetic field of B.
The latter would lead to an oscillation at 44 Hz in the radio beam
direction, due to effects either in the emission region or during
propagation out of the magnetosphere, taking the beam in and out of our
line of sight.

The predicted beat frequency varies from 0.160 to 0.205 cycles/period over
the orbit. However, the fact that we only see this drifting phenomenon at
certain orbital phases can be understood given the geometry of the system.
The energy and momentum from the 44-Hz electromagnetic radiation distort
B's magnetosphere into a cometary shape. When B is between us and A
(i.e.~the orbital phase range plotted in Fig.~3, the B
pulses propagate through the cometary tail and are relatively undisturbed.
However, for the orbital phase range plotted in Fig.~1,
the radiation from A is more transverse to the field lines in the emission
and propagation zones, allowing the modulation process to be effective
(see Fig.~4 in Lyne et al. 2004). We would expect to see this modulation
at other orbital phases where the direction of A's radiation is transverse
to the direction of B's emission. Unfortunately, however, at other orbital
phases the emission from B is too weak for these effects to be detectable.
It is not likely that this interaction is itself responsible for the
overall orbital modulation of the B-pulse emission (i.e. the large changes
in flux density and pulse profile across the orbit), although this must
also be due to the effect of A's radiation on the B magnetosphere.

In summary, we have detected drifting features in the radiation from the
2.8-s B pulsar of the double pulsar system. This effect is only detectable
at orbital phases when the electromagnetic radiation from A meets the beam
of B from the side. We have shown that this phenomenon is due to the {\it
direct influence} of the magnetic-dipole radiation from A on B. This is
clear evidence for an interaction between these two pulsars and is
extremely important for understanding, not only the unique emission
processes in this system, but also pulsar radio emission in general.
Further detailed studies of this drifting behavior may enable us to
significantly improve our knowledge of the geometry of and physics
responsible for B's emission beam.

\begin{figure*}
\begin{center}
\epsfig{file=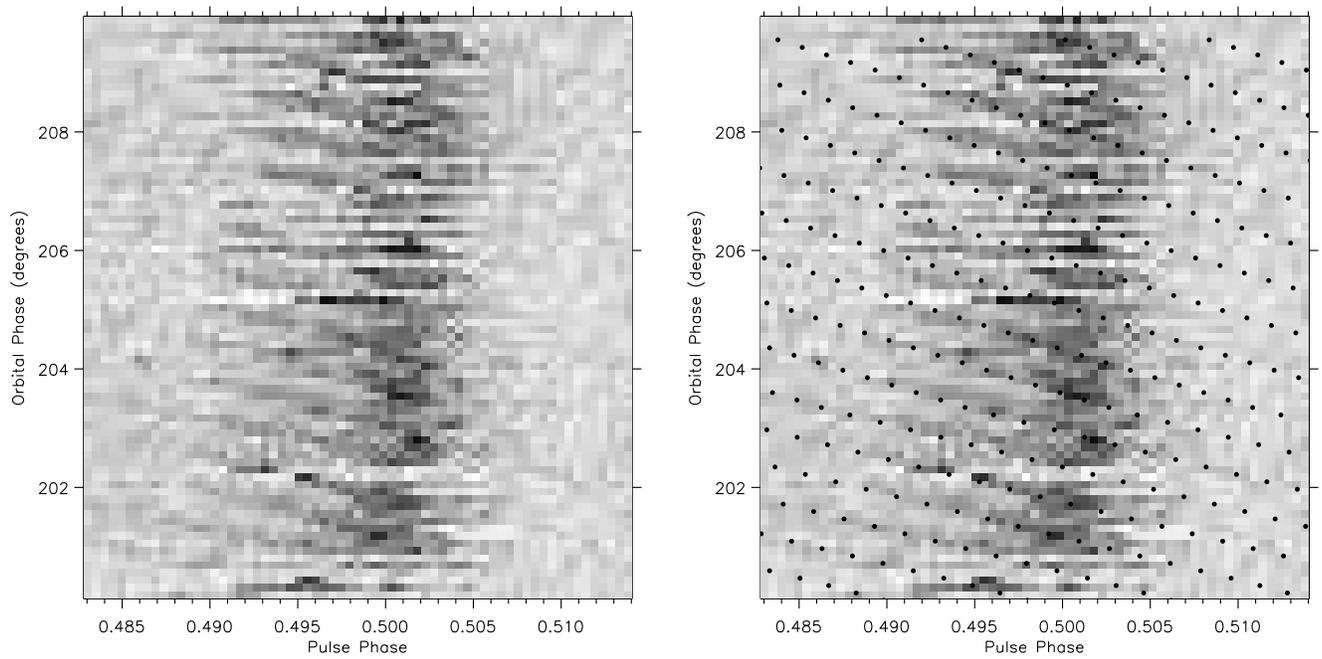,height=9cm}
\caption[]{
\label{fig:eureka}
Left:  An expanded view of Fig.~1 from orbital phases
200$\deg$ to 210$\deg$. Right: Dots denote the arrival at the centre of B of emission
from an arbitrary rotational phase of A, retarded by the
propagation time across the orbit.
}
\end{center}
\end{figure*}

\acknowledgments

We thank F. Graham Smith and A. Spitkovsky for useful discussions and N. D. R. Bhat for
writing an interface for reading the SPIGOT data. We thank the National
Radio Astronomy Observatory for making these observations publically
available.  The National Radio Astronomy Observatory is facility of the
National Science Foundation operated under cooperative agreement by
Associated Universities, Inc.  IHS holds an NSERC UFA and is supported by
a Discovery Grant. DRL is a University Research Fellow funded by the Royal
Society.  FC acknowledges support from NSF grant AST-02-05853 and a NRAO
travel grant. NDA, AP and MB received support from the Italian Ministry of
University and Research (MIUR) under the national program {\it Cofin
2003}.

{}

\end{document}